\documentclass[a4paper,fleqn]{cas-sc}

\usepackage[numbers,sort&compress]{natbib}

\usepackage{ulem}

\def\tsc#1{\csdef{#1}{\textsc{\lowercase{#1}}\xspace}}
\tsc{WGM}
\tsc{QE}

\definecolor{dkgreen}{rgb}{0.13, 0.55, 0.13}

\begin{document}
\let\WriteBookmarks\relax
\def\floatpagepagefraction{1}
\def\textpagefraction{.001}

\shorttitle{Energy-barrier characterization of secured basins in coupled oscillators with inertia}

\shortauthors{Lee et al.}

\title[mode=title]{Energy-barrier characterization of secured basins in coupled oscillators with inertia}

\author[1]{Jaiyong Lee}[orcid=0009-0005-0220-9654]
\fnmark[1]
\credit{Conceptualization, Methodology, Formal analysis, Investigation, Software, Visualization, Writing -- original draft, review \& editing}

\author[2]{Daekyung Lee}[orcid=0000-0003-2640-5658]
\fnmark[1]
\credit{Conceptualization, Methodology, Formal analysis, Investigation, Writing -- review \& editing}

\author[3]{Seung-Woo Son}[orcid=0000-0003-2244-0376]
\credit{Investigation, Methodology, Writing -- review \& editing}

\author[4]{Sang Hoon Lee}[orcid=0000-0003-3079-5679]
\credit{Investigation, Writing -- review \& editing}

\author[5]{Mi Jin Lee}[orcid=0000-0003-0013-7907]
\credit{Investigation, Writing -- review \& editing}

\author[1]{Heetae Kim}[orcid=0000-0002-5688-2626]
\cormark[1]
\ead{hkim@kentech.ac.kr}
\credit{Conceptualization, Investigation, Visualization, Supervision, Funding acquisition, Project administration, Writing -- review \& editing}

\affiliation[1]{organization={Department of Energy Engineering, Korea Institute of Energy Technology},
            city={Naju},
            postcode={58330},
            country={Republic of Korea}}

\affiliation[2]{organization={Supply Chain Intelligence Institute Austria},
            city={Vienna},
            postcode={1030},
            country={Austria}}

\affiliation[3]{organization={Department of Applied Physics, Hanyang University},
            city={Ansan},
            postcode={15588},
            country={Republic of Korea}}



\affiliation[4]{organization={Department of Physics, Gyeongsang National University},
            city={Jinju},
            postcode={52828},
            country={Republic of Korea}}

\affiliation[5]{organization={Department of Physics, Pusan National University},
            city={Busan},
            postcode={46241},
            country={Republic of Korea}}

\fntext[1]{These authors contributed equally to this work.}

\cortext[1]{Corresponding author}

\begin{abstract}
The second-order Kuramoto model captures rotor dynamics relevant to synchronization in power grids.
In actual power grids, violation of angle-stability limits can trigger protective actions that are not included in the model, restricting its validity to trajectories that remain within these limits.
We therefore define the secured basin as the region of disturbance space comprising disturbances whose trajectories remain within the angle-stability limits throughout the transient.
We then characterize the secured basin through energy barriers, quantifying their characteristic energy scale and heterogeneity across perturbation directions.
Across seven power-grid models, this characterization revealed angle-stability vulnerabilities overlooked when stability is assessed solely by the final synchronization state.
It also showed that the two energy-barrier measures characterizing the secured basin exhibit distinct associations with structural connectivity and dynamical parameters.
This characterization further offers practical advantages, as energy thresholds derived from state-space perturbations remain effective in distinguishing secured and unsecured outcomes under short-duration power disturbances and can be estimated efficiently by concentrating simulations near the secured-basin boundary.
Overall, the secured-basin framework provides a useful basis for distinguishing node-level structural and dynamical influences on transient stability in oscillator networks with synchronization constraints.
\end{abstract}



\maketitle

\section{Introduction}\label{sec:introduction}
Understanding the stability of synchronized states is a central problem in the study of coupled dynamical systems~\cite{pikovsky2001synchronization,osipov2007synchronization}.
Various approaches have been developed to characterize this stability, ranging from local analyses around a synchronized state to nonlinear measures based on finite perturbations~\cite{pecora1998master, barahona2002synchronization, hong2004factors, strogatz2024nonlinear}.
Among these approaches, nonlinear stability analysis addresses responses to large perturbations that drive the system far from its operating state and therefore requires tracking the resulting trajectories over time to determine whether synchrony is maintained or recovered.
However, evaluating such responses over a broad range of initial conditions requires repeated dynamical simulations, while both the dimension of the state space and the region to be explored grow rapidly with system size.
As a result, directly resolving the complex geometry of the basin of attraction, the region of initial conditions in state space that converge to synchrony, becomes increasingly computationally demanding.

Basin stability was introduced to reduce the computational burden of nonlinear stability assessment by avoiding reconstruction of the full basin geometry, instead randomly sampling initial perturbations in state space and evaluating the fraction that converge to synchrony~\cite{menck2013basin}.
Owing to this computational advantage, basin stability has been widely applied to stability assessment in power grids~\cite{schultz2014detours, menck2014dead, kim2015community, kim2016building, kim2018multistability, kim2019structural, witthaut2022collective}, as well as in various other complex networks~\cite{anderies2013topology, mirollo1990synchronization, tass2007phase}.
Because basin stability primarily evaluates whether the modeled physical system ultimately recovers synchrony after a perturbation, it does not account for whether the transient trajectory remains within constraints imposed by the physical environment.
This limitation has motivated the view that transient stability should also account for whether trajectories remain within prescribed admissible regions of state space over a finite observation period, irrespective of the system's final state~\cite{hellmann2016survivability,nitzbon2017deciphering}.

This perspective becomes particularly important in power grids, which can be modeled as networks of coupled oscillators with inertia, such as the second-order Kuramoto model.
In practical power-system operation, protective actions depend on phase-angle differences between synchronous machines, which are directly related to angle stability, a key measure of power-system stability, and can trigger changes in network topology or component operating states when allowable limits are exceeded~\cite{kundur2004definition,machowski2020power}.
Consequently, even if a trajectory of an oscillator model representing a power grid is eventually observed to recover synchrony, the model trajectory cannot readily be interpreted as a physically realizable representation of the actual grid response if it includes states beyond a phase-angle limit at which protective intervention would occur.

To address the characteristic angle-stability constraints of power grids, we propose a node-wise stability assessment framework that considers not only whether a perturbed network recovers synchrony but also whether its transient trajectory remains within the angle-stability constraints required for physically meaningful grid operation, while reducing the dependence of existing measures on an analyst-specified perturbation range.
We define the secured basin as the region of perturbations in the state space of an individual node whose trajectories remain within the phase-angle limits intrinsic to power-grid operation throughout the transient, thereby avoiding protective actions that may alter the network topology or nodal power injections.
We further characterize the secured-basin boundary as an energy barrier and quantify it using two measures for evaluating robustness against perturbations: the mean barrier height and the barrier height disparity.

Using power-grid datasets from multiple countries, we examine the utility of the proposed secured basin and the energy-barrier characteristics of its boundary beyond existing stability measures~\cite{lee2024reinforcement,pagnier2019inertia,tyloo2019key,song2025kpg}.
First, we examine whether distinguishing the secured basin from the synchronized basin reveals aspects of robustness that are not captured by existing measures, and investigate the physical meaning of the mean barrier height and barrier height disparity derived from the secured-basin boundary.
We then evaluate whether the energy characteristics obtained in the state-perturbation space retain their validity when disturbances are represented in a space more directly related to power-grid fault assessment, and whether these characteristics can be estimated efficiently.

\section{Concepts}\label{sec:safe}

\subsection{Secured basin}\label{subsec:basin_safety}

The second-order Kuramoto model describes the phase dynamics of a network of coupled oscillators with inertia and damping.
We consider a network of \(N\) oscillators, where each node \(i\) represents an oscillator.
Let \(\theta_i\) and \(\omega_i=\dot{\theta}_i\) denote the phase angle and angular velocity of node \(i\), respectively.
The dynamics are governed by
\begin{equation}
\dot{\theta}_i=\omega_i,
\label{eq:theta_dot_concept}
\end{equation}
\begin{equation}
M_i\dot{\omega}_i
=
P_i
-
D_i\omega_i
+
\sum_j K_{ij}\sin(\theta_j-\theta_i),
\label{eq:swing_concept}
\end{equation}
where \(M_i\), \(D_i\), \(P_i\), and \(K_{ij}\) denote the inertia, damping coefficient, net power injection, and coupling strength, respectively. (see Sec.~\ref{subsec:methods_model}).

Within this framework, the stability of each node can be assessed by perturbing the phase angle and angular velocity of a specific node from the steady state and then determining whether the trajectory initiated from that initial condition returns to synchrony.
However, even if the trajectory predicted by the model eventually reaches a synchronized state, once it crosses a secured operating limit (defined below) at which protective actions would intervene in the actual power grid, the subsequent model trajectory can no longer be readily interpreted as a physically representative response of the original grid.
We therefore define the secured basin of a target node as the set of initial perturbations to its phase angle and angular velocity whose resulting trajectories remain within the secured operating region throughout the transient.
Initial perturbations outside the secured basin are classified as unsecured because their resulting trajectories violate the prescribed phase-angle limit, irrespective of whether synchrony is eventually recovered.

We consider a network \(G=(\mathcal{V},\mathcal{E})\), where \(\mathcal{V}\) and \(\mathcal{E}\) denote the sets of nodes and edges, respectively.
The secured operating region in state space is defined as
\begin{equation}
X_{\mathrm{secured}}
=
\left\{
\mathbf{x}
=
\bigl(\boldsymbol{\theta},\boldsymbol{\omega}\bigr)
:
\left|
\bigl(\theta_i-\theta_j\bigr)
\bmod 2\pi
\right|
<
\Delta\theta_{\mathrm{crit}}
\quad
\forall(i,j)\in\mathcal{E}
\right\}.
\label{eq:X_secured}
\end{equation}
Here, \(\boldsymbol{\theta}\) and \(\boldsymbol{\omega}\) denote the vectors of nodal phase angles and angular velocities, respectively.
The symmetric modulo operation, \(x\bmod 2\pi\in(-\pi,\pi]\), maps a phase-angle difference to its principal interval.

The choice of \(\Delta\theta_{\mathrm{crit}}=\pi\) is consistent with the interpretation of large rotor-angle separation in power-system transient-stability assessment, where rotor-angle differences approaching or exceeding \(180^\circ\) are associated with loss of synchronism~\cite{kundur2007power, sobbouhi2020online, sobbouhi2021transient}.
Following this power-system interpretation, we set \(\Delta\theta_{\mathrm{crit}}=\pi\) as the phase-angle limit of the secured operating region and exclude trajectories that reach this limit from the secured region.
Mathematically, some trajectories can still recover synchrony in the model even when their phase-angle evolution extends beyond this limit, but they are classified as unsecured to reflect the practical constraint that such large angular excursions may activate protective actions in actual power systems.

For a target node \(i\), let $\Delta\mathbf{x}_i=(\Delta\theta_i,\Delta\omega_i)$ denote a perturbation applied to node $i$.
The corresponding full-system initial state is
\[
\mathbf{x}^{(i)}(t=0;\Delta\mathbf{x}_i)
=
\left(
\mathbf{x}_1^{\ast},\ldots,
\mathbf{x}_{i}^{\ast}+\Delta\mathbf{x}_i,\ldots,
\mathbf{x}_N^{\ast}
\right),
\]
where $\mathbf{x}_{i}^{\ast}=(\theta_i^{\ast},0)$ denotes the synchronized fixed-point state of node $i$.
The superscript $(i)$ indicates that the full-system state results from a perturbation applied to node $i$.

The resulting full-system trajectory over the observation horizon  $T$ is
\begin{equation}
\Gamma^{(i)}(\Delta\mathbf{x}_i)
=
\left\{
\mathbf{x}^{(i)}(t;\Delta\mathbf{x}_i)
\mid
t\in[0,T]
\right\},
\label{eq:trajectory}
\end{equation}
where \(\mathbf{x}^{(i)}(t;\Delta\mathbf{x}_i)\) denotes the full-system state at time \(t\) evolving from the initial condition \(\mathbf{x}^{(i)}(t=0;\Delta\mathbf{x}_i)\). 
A perturbation is classified as secured if its entire trajectory remains within the secured operating region,
\[
\Gamma^{(i)}(\Delta\mathbf{x}_i)\subseteq X_{\mathrm{secured}}.
\]
Equivalently, \(\Gamma^{(i)}(\Delta\mathbf{x}_i)\subseteq X_{\mathrm{secured}}\) requires that no connected pair of nodes crosses the \(\pi\) phase-angle boundary at any time during the observation horizon.

For a target node $i$, let $A_i$ denote the perturbation set for node $i$.
The secured perturbation set \(\mathcal{S}_i\) is then defined as
\begin{equation}
\mathcal{S}_i
=
\left\{
\mathbf{\Delta x}_{i}\in A_i :
\Gamma^{(i)}(\Delta\mathbf{x}_i)
\subseteq X_{\mathrm{secured}}
\right\},
\label{eq:S_i}
\end{equation}
and the unsecured perturbation set \(\mathcal{U}_i\) is defined as
\begin{equation}
\mathcal{U}_i
=
A_i \setminus \mathcal{S}_i~,
\label{eq:U_i}
\end{equation}
where $\setminus$ indicates the set difference.

\begin{figure}
    \centering
    \includegraphics[width=0.8\textwidth]{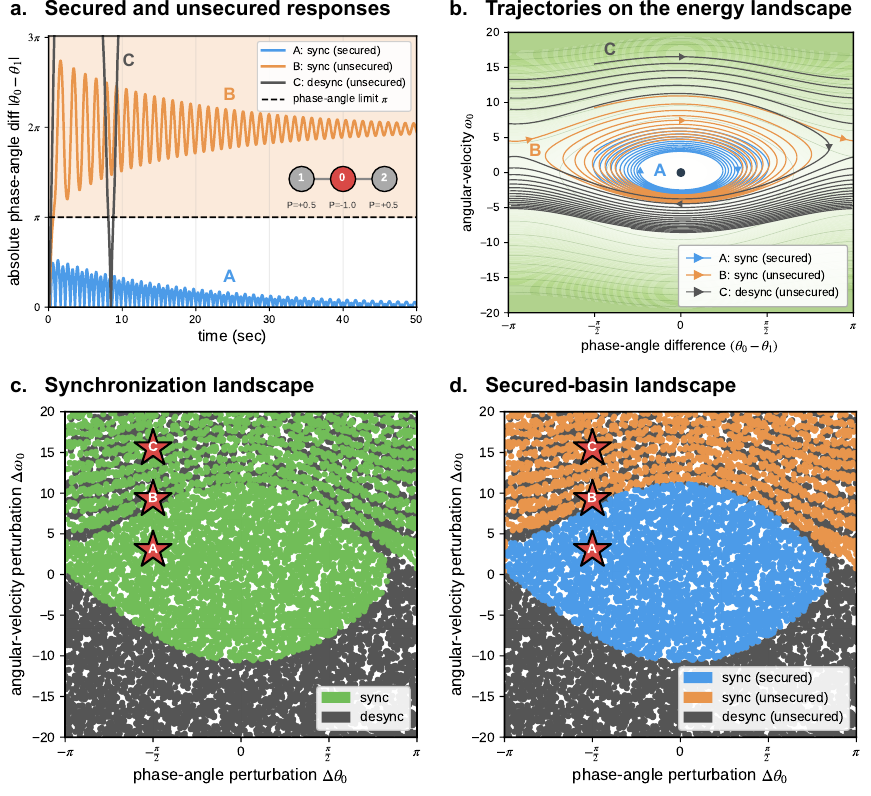}
    \caption{
    Classification of perturbation scenarios according to synchronization recovery and transient phase-angle constraints for perturbations applied to node 0.
    (a) Time evolution of the absolute phase-angle difference
    \(
    |\theta_0(t)-\theta_1(t)|
    \)
    for trajectories generated by three perturbations,
    \(\Delta \mathbf{x}_0^{A}\),
    \(\Delta \mathbf{x}_0^{B}\),
    \(\Delta \mathbf{x}_0^{C}\in A_0\),
    in a three-node chain oscillator network governed by second-order Kuramoto dynamics.
    The network has the topology \(1-0-2\), with \(K=10\), \(P_0=-1.0\), and \(P_1=P_2=0.5\).
    The trajectories \(\Gamma^{(0)}(\Delta\mathbf{x}_0^{A})\) and \(\Gamma^{(0)}(\Delta\mathbf{x}_0^{B})\) recover the synchronized operating state,
    but only the trajectory \(\Gamma^{(0)}(\Delta\mathbf{x}_0^{A})\) remains within $X_{\mathrm{secured}}$ throughout the observation horizon.
    The trajectory \(\Gamma^{(0)}(\Delta\mathbf{x}_0^{C})\) fails to recover synchrony.
    (b) Schematic representation of the same trajectories on an energy landscape.
    The three initial conditions produce a secured outcome, an unsecured synchronized outcome, and an unsecured desynchronized outcome, respectively.
    (c) Perturbation set \(A_0\) of node 0,
    where each perturbation
    ${\Delta\mathbf{x}_0}=(\Delta\theta_0,\Delta\omega_0)\in A_0$
    is classified according to synchronization recovery.
    Basin stability of node $0$ is \(p_0^{\mathrm{BS}}=0.534\).
    (d) The same perturbation set classified according to whether the resulting trajectory remains within $X_{\mathrm{secured}}$ throughout the observation horizon.
    The relative secured-basin volume of node $0$ is \(p_0^{\mathrm{S}}=0.346\).
    Here, \(X_{\mathrm{secured}}\) is defined using
    \(\Delta\theta_{\mathrm{crit}}=\pi\), and all trajectories are integrated up to \(T=500\).
    }
    \label{fig:basin_safety}
\end{figure}

The distinction is illustrated using the simple three-oscillator network $G_3=(\mathcal{V}_3,\mathcal{E}_3)$ governed by second-order Kuramoto dynamics defined in Eqs.~\ref{eq:theta_dot_concept} and ~\ref{eq:swing_concept}, as shown in Fig.~\ref{fig:basin_safety}a, with $|\mathcal{V}_3|=3$ and $|\mathcal{E}_3|=2$, where $|\cdots|$ indicates a cardinality of a given set.
The three trajectories originate from three distinct initial conditions,
\(\Delta \mathbf{x}_0^{A}\),
\(\Delta \mathbf{x}_0^{B}\), and
\(\Delta \mathbf{x}_0^{C}\),
selected from the perturbation set $A_0$ for the target node $0$.
The detailed settings for the state perturbations and outcome classification are provided in Sec.~\ref{subsec:methods_state_perturbation}.
The trajectories \(\Gamma^{(0)}(\Delta\mathbf{x}_0^{A})\) and \(\Gamma^{(0)}(\Delta\mathbf{x}_0^{B})\) both recover the synchronized state, but only the former remains within the prescribed phase-angle limit throughout the transient.
By contrast, the trajectory \(\Gamma^{(0)}(\Delta\mathbf{x}_0^{C})\) fails to recover synchrony.
Fig.~\ref{fig:basin_safety}b represents the same three cases on an energy landscape, distinguishing a secured outcome, an unsecured synchronized outcome, and an unsecured desynchronized outcome.

For the same perturbation set \(A_i\), basin stability is given by the fraction of initial perturbations whose trajectories eventually recover the synchronized state~\cite{menck2013basin}:
\begin{equation}
p_i^{\mathrm{BS}}
=
\frac{\mathrm{Vol}(\mathcal{B}_i)}
{\mathrm{Vol}(A_i)},
\label{eq:basin_stability}
\end{equation}
where \(\mathcal{B}_i\) denotes the set of initial perturbations that return to synchrony, and $\mathrm{Vol}(\cdot)$ denotes the measure of a set in
the perturbation space. In the two-dimensional single-node perturbation
space considered here, this measure corresponds to its area.
For comparison with the basin stability $p_i^{\mathrm{BS}}$, we define the relative secured-basin volume from the secured perturbation set $\mathcal{S}_i$ in Eq.~\ref{eq:S_i} as its volume fraction within the perturbation set \(A_i\):
\begin{equation}
p_i^{\mathrm{S}}
=
\frac{\mathrm{Vol}(\mathcal{S}_i)}
{\mathrm{Vol}(A_i)}.
\label{eq:volume}
\end{equation}
Under uniform sampling, this volume fraction can be approximated by the fraction of sampled perturbations belonging to \(\mathcal{S}_i\).

Fig.~\ref{fig:basin_safety}c,d compare these two classifications of the same single-node perturbation plane.
Basin stability \(p_i^{\mathrm{BS}}\) classifies perturbations solely according to whether synchrony is eventually restored, whereas \(p_i^{\mathrm{S}}\) additionally requires the entire transient trajectory to remain within \(X_{\mathrm{secured}}\).
Consequently, some perturbations counted as synchronized recoveries by basin stability are classified as unsecured because their trajectories cross the prescribed phase-angle limit before returning to synchrony.

\subsection{Energy-barrier characterization}
\label{subsec:energy_safety}

The secured basin obtained for each node provides a clearly identifiable boundary in perturbation space, as illustrated in Fig.~\ref{fig:basin_safety}d, unlike the synchronization landscape in Fig.~\ref{fig:basin_safety}c, where such a boundary is difficult to define.
This secured-basin boundary therefore provides a natural basis for quantifying node-level stability in terms of how easily or difficultly different perturbations drive the system out of the secured basin.
In the following, we first characterize the energy required to reach this boundary as an energy barrier, then summarize the distribution of barriers arising from different perturbation conditions using two measures, and finally introduce an efficient strategy for estimating them.

\subsubsection{Energy barrier}
\label{subsubsec:directional_energy_barriers}

The relative secured-basin volume \(p_i^{\mathrm{S}}\) in Eq.~\ref{eq:volume} enables a direct comparison with basin stability \(p_i^{\mathrm{BS}}\)~\cite{menck2013basin} in Eq.~\ref{eq:basin_stability}, thereby revealing the difference in stability that arises when the secured operating condition is imposed during the transient.
However, because \(p_i^{\mathrm{S}}\) is defined as the fraction of the secured basin within a perturbation region specified by the analyst for sampling, its value can be substantially affected by the chosen perturbation range.
The full range of phase-angle perturbation \(\Delta\theta_i\) can be specified relatively clearly as \([-\pi,\pi]\) because of phase periodicity, whereas the range of angular-velocity perturbation \(\Delta\omega_i\) is set to \([-\omega_{\max},\omega_{\max}]\).
Unlike the phase angle, no canonical value of \(\omega_{\max}\) exists.
Consequently, the selected range of angular-velocity perturbations directly affects the relative volume and the resulting stability assessment.

Moreover, systems with similar \(p_i^{\mathrm{S}}\) may withstand perturbations of substantially different physical magnitudes.
This is because \(p_i^{\mathrm{S}}\), as a relative fraction, does not provide a common physical scale for the magnitude of perturbations that drive the system out of the secured operating region.
Furthermore, reducing the outcomes within a perturbation set to a single relative-volume statistic can discard important stability information contained in the secured basin.

Instead of treating all sampled perturbations as equivalent contributions to stability, as in basin-stability analysis, we consider the perturbation energy to retain the physical magnitude of each perturbation.
The perturbation energy applied to node \(i\) is defined as the sum of the kinetic-energy change induced by the initial angular-velocity perturbation and the potential-energy change induced by the initial phase-angle perturbation,
\begin{equation}
E_i(\Delta\mathbf{x}_i)
=
\Delta T_i(\Delta\omega_i)
+
\Delta V_i(\Delta\theta_i).
\label{eq:perturbation_energy}
\end{equation}
This energy representation provides a common physical scale for comparing perturbations with different combinations of phase-angle and angular-velocity changes.
The detailed calculation of \(E_i\) for the second-order Kuramoto model is provided in Sec.~\ref{subsec:methods_energy}.

Each perturbation is characterized not only by its energy but also by its direction in the \((\Delta\theta_i,\Delta\omega_i)\) perturbation space.
Because the phase-angle and angular-velocity perturbations have different physical dimensions, we define the perturbation direction in a normalized coordinate system,
\begin{equation}
\tilde{\mathbf{x}}_i
=
\left(
\frac{\Delta\theta_i}{\pi},
\frac{\Delta\omega_i}{\omega_{\max}}
\right),
\end{equation}
where \(\pi\) and \(\omega_{\max}\) are used as normalization scales for the phase-angle and angular-velocity perturbations, respectively.
For a perturbation \(\Delta\mathbf{x}_i=(\Delta\theta_i,\Delta\omega_i)\in A_i\), the perturbation direction is then defined by
\begin{equation}
\hat{\mathbf d}_i
=
\frac{\tilde{\mathbf{x}}_i}
{\left|\tilde{\mathbf{x}}_i\right|}.
\label{eq:direction}
\end{equation}
Along different normalized perturbation directions, the secured-basin boundary can be reached at different perturbation energies.
The perturbation energy at the boundary along direction \(\hat{\mathbf d}_i\) defines the directional energy barrier \(E_i^{\mathrm{B}}(\hat{\mathbf d}_i)\), which represents on a common physical energy scale how much perturbation the secured basin can withstand in that direction.


\subsubsection{Energy thresholds and barrier measures}
\label{subsubsec:energy_thresholds_barrier_measures}

Although the directional energy barriers \(E_i^{\mathrm{B}}(\hat{\mathbf d}_i)\) retain detailed information on the secured-basin geometry, our aim is to compress the information distributed across many perturbation directions into a small number of physically interpretable quantities while preserving its essential features.
From a power-grid perspective, quantities of practical interest are the upper energy limit below which all perturbations remain secured and the lower energy limit above which all perturbations are consistently unsecured.
These two energy scales correspond to the minimum and maximum of the directional energy barriers \(E_i^{\mathrm{B}}(\hat{\mathbf d}_i)\) that constitute the secured-basin boundary.
Rather than estimating the barrier separately along every perturbation direction at high computational cost, we calculate the perturbation energies of the sampled perturbations in \(A_i\), classify their outcomes as secured or unsecured, and extract the two energy thresholds from the extrema of the resulting outcome sets.
For each perturbation condition \(\Delta\mathbf{x}_i\in A_i\), the corresponding perturbation energy \(E_i(\Delta\mathbf{x}_i)\) is calculated from Eq.~\ref{eq:perturbation_energy}.

Using the secured and unsecured perturbation sets in Eqs.~\ref{eq:S_i} and \ref{eq:U_i}, we define two direction-independent energy thresholds for each node.
The secured energy threshold is defined as the lowest energy among the unsecured outcomes,
\begin{equation}
E_i^{\mathrm{S}}
=
\min_{\Delta\mathbf{x}_i \in \mathcal{U}_i}
E_i(\Delta\mathbf{x}_i),
\label{eq:secured_energy}
\end{equation}
whereas the unsecured energy threshold is defined as the highest energy among the secured outcomes,
\begin{equation}
E_i^{\mathrm{U}}
=
\max_{\Delta\mathbf{x}_i \in \mathcal{S}_i}
E_i(\Delta\mathbf{x}_i).
\label{eq:unsecured_energy}
\end{equation}
It is important to note that \(E_i^{\mathrm{S}}\) and \(E_i^{\mathrm{U}}\) are named according to the energy regimes they delimit, rather than according to the outcome class from which each threshold is extracted.
Accordingly, all sampled perturbations with energies below \(E_i^{\mathrm{S}}\) are secured, whereas all sampled perturbations with energies above \(E_i^{\mathrm{U}}\) are unsecured.
Between \(E_i^{\mathrm{S}}\) and \(E_i^{\mathrm{U}}\), secured and unsecured outcomes coexist, meaning that perturbation energy alone, without information on perturbation direction, is insufficient to determine whether a trajectory remains within the secured operating region.
We therefore refer to this range as the mixed-outcome energy interval,
\begin{equation}
\mathcal{M}_i
=
\left[
E_i^{\mathrm{S}},
E_i^{\mathrm{U}}
\right].
\label{eq:interval}
\end{equation}

The potential energy \(\Delta V_i(\Delta\theta_i)\) in Eq.~\ref{eq:perturbation_energy} is evaluated relative to the reference stable equilibrium \(\boldsymbol{\theta}^{\ast}\).
As the phase angle moves away from this equilibrium toward a neighboring unstable equilibrium, \(\Delta V_i(\Delta\theta_i)\) increases and reaches a local maximum at the unstable-equilibrium angle.
Beyond this angle, \(\Delta V_i(\Delta\theta_i)\) decreases again even though the perturbed state has already crossed the corresponding energy barrier.
Consequently, a perturbation located beyond the barrier can have a lower energy than the barrier itself.
This creates an ambiguity in our threshold-extraction procedure because taking the minimum perturbation energy from the unsecured set to identify the lowest energy barrier does not necessarily select a point on the secured-basin boundary.
To avoid this ambiguity, the energy-barrier analysis is restricted to the portion of the principal phase-perturbation interval \(\Delta\theta_i\in(-\pi,\pi]\) that does not extend beyond the local unstable equilibrium, and only perturbations within this range are used to evaluate the energy-based measures.
The detailed energy definition and treatment of the unstable-equilibrium bound are provided in Sec.~\ref{subsec:methods_energy}.

\begin{figure}
    \centering
    \includegraphics[width=0.8\textwidth]{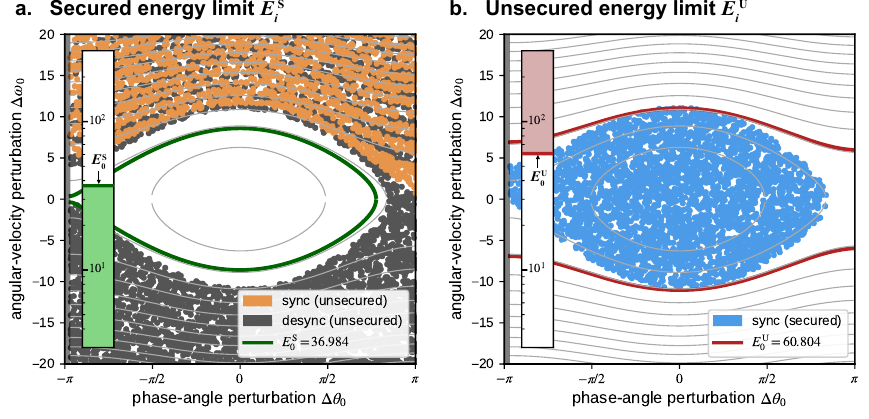}
    \caption{
    Energy thresholds identified in the perturbation plane $A_0$ of node \(0\) for the three-node oscillator system.
    (a) Unsecured outcomes.
    Orange points represent unsecured synchronized outcomes that return to synchrony after crossing the allowable phase-angle limit, whereas gray points represent unsecured desynchronized outcomes that fail to recover synchrony.
    The green contour is the constant-energy contour passing through the lowest-energy unsecured outcome and defines
    \(E_0^{\mathrm{S}}=36.984\).
    (b) Secured outcomes in the same perturbation plane.
    Blue points represent trajectories that return to synchrony while remaining within the secured operating region throughout the transient.
    The red contour is the constant-energy contour passing through the highest-energy secured outcome and defines
    \(E_0^{\mathrm{U}}=60.804\).
    The resulting mixed-outcome energy interval is $\mathcal{M}_0=[36.984,60.804]$, with mean barrier height $\overline{E}_0=47.421$ and barrier height disparity $\Delta_0=1.282$.
    The inset color bars show that all perturbations below $E_0^{\mathrm{S}}$ are secured and shaded in light green, 
    whereas all perturbations above $E_0^{\mathrm{U}}$ are unsecured and shaded in light red.
    Thin gray curves in both panels indicate additional contours of constant perturbation energy level.
    The black-shaded region denotes perturbations outside the unstable-equilibrium bounds, which are excluded from the energy analysis.
    }
    \label{fig:energy_safety}
\end{figure}

Fig.~\ref{fig:energy_safety} illustrates how the two energy thresholds are determined for the same three-node oscillator system introduced in Sec~\ref{subsec:basin_safety}.
In Fig.~\ref{fig:energy_safety}a, the unsecured trajectory with the lowest energy determines \(E_0^{\mathrm{S}}\), below which no unsecured outcome is observed.
In contrast, in Fig.~\ref{fig:energy_safety}b, the secured trajectory with the highest energy determines \(E_0^{\mathrm{U}}\), above which no secured outcome is observed.
For perturbations with energies between \(E_0^{\mathrm{S}}\) and \(E_0^{\mathrm{U}}\), both secured and unsecured outcomes are observed in the sampled data, so their outcomes cannot be determined from perturbation energy alone without information on perturbation direction.
This range therefore corresponds to the mixed-outcome energy interval \(\mathcal{M}_0\).

Because the secured-basin boundary can be reached at different energies \(E_i^{\mathrm B}(\hat{\mathbf d}_i)\) along different perturbation directions, the directional energy-barrier height can vary with \(\hat{\mathbf d}_i\).
Although \(E_i^{\mathrm S}\) and \(E_i^{\mathrm U}\), estimated from the sampled perturbations, characterize the lower and upper extents of \(E_i^{\mathrm B}(\hat{\mathbf d}_i)\), we further introduce two measures, the mean barrier height \(\overline{E}_i\) and the barrier height disparity \(\Delta_i\), to describe the node-level energy-barrier landscape in terms of its characteristic energy scale and directional heterogeneity, respectively.
We adopt a multiplicative (geometric) representation, in which the separation between \(E_i^{\mathrm S}\) and \(E_i^{\mathrm U}\) is evaluated relative to the characteristic energy scale rather than by their absolute difference.
This relative representation allows nodes with different characteristic energy scales to be compared consistently, as the same absolute separation between the two thresholds can be interpreted as substantially different levels of barrier heterogeneity depending on the barrier energy scale of the node.

For node $i$, the mean barrier height $\overline{E}_i$ is defined as
\begin{equation}
\overline{E}_i
=
\sqrt{
E_i^{\mathrm{S}}
\times
E_i^{\mathrm{U}}
},
\label{eq:geometric_mean}
\end{equation}
and the barrier height disparity $\Delta_i$ is defined as
\begin{equation}
\Delta_i
=
\sqrt{
\frac{
E_i^{\mathrm{U}}
}{
E_i^{\mathrm{S}}
}
\label{eq:disparity}
}.
\end{equation}
Here, $\overline{E}_i$ represents the characteristic energy scale, while $\Delta_i$ quantifies directional heterogeneity of two energy thresholds $E_i^{\mathrm{S}}$ and $E_i^{\mathrm{U}}$.

Using the mean barrier height $\overline{E}_i$ and barrier height disparity $\Delta_i$ defined in Eqs.~\ref{eq:geometric_mean} and \ref{eq:disparity}, the mixed-outcome energy interval $\mathcal{M}_i$ defined in Eq.~\ref{eq:interval} can be interpreted as
\[
\mathcal{M}_i
=
\left[
E_i^{\mathrm{S}},
\,
E_i^{\mathrm{U}}
\right]
=
\left[
\frac{\overline{E}_i}{\Delta_i},
\,
\overline{E}_i \times \Delta_i
\right].
\]
Thus, the secured and unsecured energy thresholds are symmetrically located around \(\overline{E}_i\) on a logarithmic energy scale, with \(\Delta_i\) determining their relative separation.
When \(\Delta_i=1\), the two thresholds coincide, so \(\overline{E}_i\) alone fully characterizes the energy boundary between secured and unsecured outcomes.
As \(\Delta_i\) increases, the mixed-outcome energy interval broadens, reducing the representativeness of \(\overline{E}_i\) as a single characteristic scale and increasing the importance of \(\Delta_i\) in describing the energy-barrier landscape.

\subsubsection{Boundary-focused search}
\label{subsubsec:efficient_threshold_identification}

Basin stability \(p_i^{\mathrm{BS}}\) and the relative secured-basin volume \(p_i^{\mathrm{S}}\) in Eqs.~\ref{eq:basin_stability} and \ref{eq:volume}, respectively, are volume-based measures.
Their estimation therefore requires sufficiently uniform sampling over the prescribed perturbation domain so that the fraction of sampled outcomes represents the corresponding volume fraction.
In contrast, the proposed energy-barrier characteristics, the mean barrier height \(\overline{E}_i\) and barrier height disparity \(\Delta_i\), are determined only by the two energy thresholds \(E_i^{\mathrm{S}}\) and \(E_i^{\mathrm{U}}\).
Because these thresholds depend on the lowest-energy unsecured and highest-energy secured outcomes, their estimation does not require uniform coverage of the entire perturbation domain.
This allows simulations to be concentrated near energy regions capable of updating the current threshold estimates, forming the basis of the boundary-focused search.

\begin{figure}
    \centering
    \includegraphics[width=0.8\textwidth]{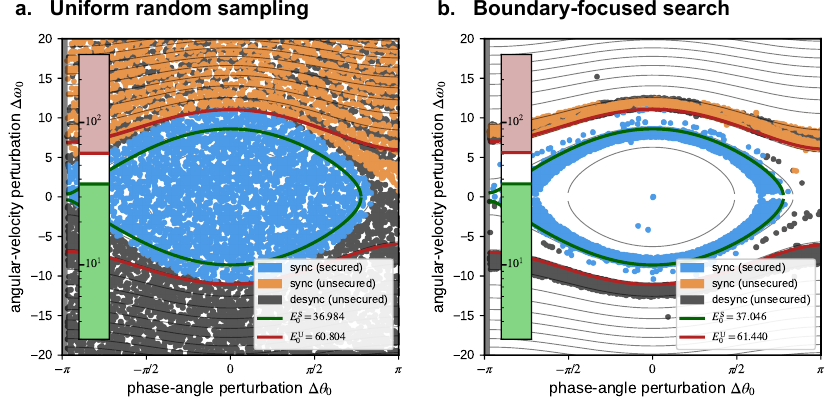}
    \caption{
    Comparison of uniform random sampling and the boundary-focused search for estimating the two energy thresholds.
    (a) Uniform random sampling in the perturbation plane of node \(0\) in the three-node chain network used throughout this section.
    Perturbations are distributed throughout the prescribed ranges of phase-angle and angular-velocity perturbations.
    (b) Boundary-focused search for the same node and system.
    After the temporary thresholds are established, sampling is concentrated in regions that can update either \(E_0^{\mathrm{S}}\) or \(E_0^{\mathrm{U}}\), while regions that cannot modify the current threshold estimates are sampled less frequently.
    In both panels, blue points indicate secured outcomes, orange points indicate unsecured synchronized outcomes, and gray points indicate unsecured desynchronized outcomes.
    The green and red curves denote the estimated secured and unsecured energy thresholds, \(E_0^{\mathrm{S}}\) and \(E_0^{\mathrm{U}}\), respectively, and the thin gray curves denote contours of constant perturbation energy.
    The black-shaded region denotes perturbations outside the unstable-equilibrium bounds, which are excluded from the energy analysis.
    The vertical energy bar indicates the secured region below \(E_0^{\mathrm{S}}\) and the unsecured region above \(E_0^{\mathrm{U}}\) on a logarithmic energy scale.
    }
    \label{fig:concept_fast}
\end{figure}

The boundary-focused search progressively updates \(E_i^{\mathrm{S}}\) and \(E_i^{\mathrm{U}}\) while preferentially sampling perturbations in energy regions that can modify either threshold.
This avoids spending simulations on large portions of the perturbation domain that do not contribute to threshold estimation and reduces the dependence on exhaustive sampling over an analyst-specified perturbation region.
Fig.~\ref{fig:concept_fast} illustrates the resulting difference in sampling patterns between uniform random sampling and the boundary-focused search.
The detailed threshold-update and sampling procedures are provided in Sec.~\ref{subsec:methods_search}.

\section{Results}
\label{sec:results}

\subsection{Robustness information beyond volume-based stability measures}
\label{subsec:high_stability}

\begin{figure}
    \centering
    \includegraphics[width=0.8\textwidth]{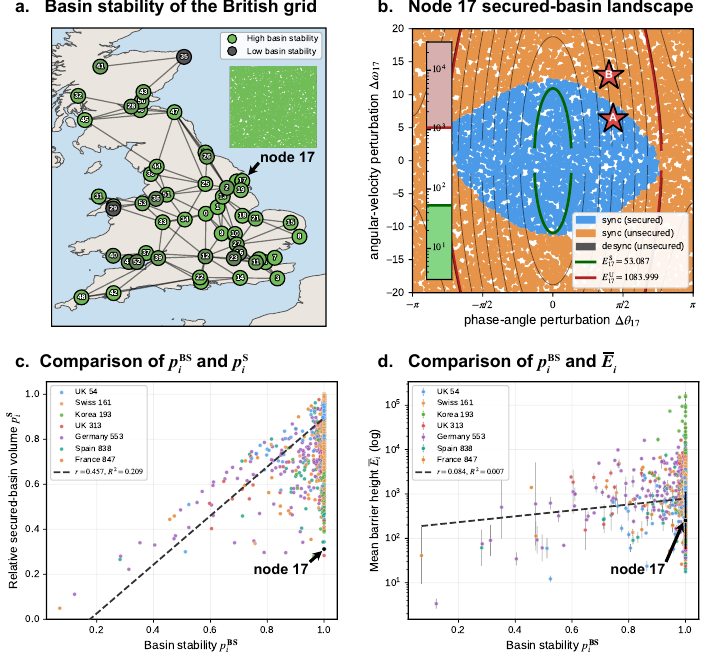}
    \caption{
    Comparison of basin stability \(p_i^{\mathrm{BS}}\) with the relative secured-basin volume \(p_i^{\mathrm{S}}\), mean barrier height \(\overline{E}_i\), and barrier height disparity \(\Delta_i\), highlighting distinct aspects of nodal robustness.
    (a) Basin stability \(p_i^{\mathrm{BS}}\) in the 54-node British power grid.
    Node color indicates \(p_i^{\mathrm{BS}}\), and the inset shows the perturbation plane and set \(A_{17}\) of node \(17\), for which \(p_{17}^{\mathrm{BS}}=1\).
    (b) Secured-basin landscape of node \(17\).
    Blue points denote secured outcomes, orange points denote unsecured synchronized outcomes, and unsecured desynchronized outcomes are absent for this node.
    The relative secured-basin volume is \(p_{17}^{\mathrm{S}}=0.308\).
    The green and red contours denote the secured and unsecured energy thresholds, \(E_{17}^{\mathrm{S}}\) and \(E_{17}^{\mathrm{U}}\), respectively.
    These thresholds give a mean barrier height of \(\overline{E}_{17}=239.880\) and a barrier height disparity of \(\Delta_{17}=4.519\).
    The perturbations \(\Delta\mathbf{x}_{17}^{A}\) and \(\Delta\mathbf{x}_{17}^{B}\) yield the same perturbation energy,
    \(E_{17}(\Delta\mathbf{x}_{17}^{A})=E_{17}(\Delta\mathbf{x}_{17}^{B})=500\),
    but lead to different security outcomes.
    (c) Node-wise relationship between basin stability \(p_i^{\mathrm{BS}}\) and relative secured-basin volume \(p_i^{\mathrm{S}}\) across the seven power-grid models.
    Each point represents one node.
    The Pearson correlation coefficient and coefficient of determination are \(r=0.457\) and \(R^2=0.209\), respectively.
    Node \(17\) of the 54-node British power grid is highlighted by a bold black point for \(p_i^{\mathrm{S}}\).
    (d) Node-wise relationship between basin stability \(p_i^{\mathrm{BS}}\) and mean barrier height \(\overline{E}_i\).
    The vertical line for each node spans the mixed-outcome energy interval \(\mathcal{M}_i=[E_i^{\mathrm{S}},E_i^{\mathrm{U}}]\), whose relative width is quantified by the barrier height disparity \(\Delta_i\).
    The Pearson correlation coefficient and coefficient of determination, calculated using the log-transformed mean barrier heights, are \(r=0.084\) and \(R^2=0.007\), respectively.
    Node \(17\) of the 54-node British power grid is highlighted by a bold black point for \(\overline{E}_{17}\) and a bold black vertical line for \(\Delta_{17}\), respectively.
    }
    \label{fig:stability_safety}
\end{figure}

We analyze seven empirical power-grid models~\cite{lee2024reinforcement,tyloo2019key,pagnier2019inertia,song2025kpg}, beginning with the reduced 54-node British power grid as a representative example.
Fig.~\ref{fig:stability_safety}a shows the basin stability $p^{\rm BS}$~\cite{menck2013basin} of each node in the 54-node British power grid.
Many nodes exhibit high basin stability, indicating a high probability of recovering synchrony in response to perturbations contained in their respective $A_i$.
Node \(17\) provides an example of high basin stability, with \(p_{17}^{\mathrm{BS}}=1\), meaning that synchrony is recovered for all perturbations in \(A_{17}\), which corresponds to maximal robustness according to basin stability within the considered perturbation domain.

Fig.~\ref{fig:stability_safety}b reveals the vulnerability that is hidden by the unit basin stability of node \(17\).
Despite \(p_{17}^{\mathrm{BS}}=1\), the relative secured-basin volume is only \(p_{17}^{\mathrm{S}}=0.308\), indicating that a large fraction of perturbations violate the transient phase-angle constraint even though synchrony is eventually recovered.
The two energy thresholds are also widely separated, with \(E_{17}^{\mathrm{S}}=53.087\) and \(E_{17}^{\mathrm{U}}=1083.999\), yielding a mean barrier height of \(\overline{E}_{17}=239.880\) and a relatively large barrier height disparity of \(\Delta_{17}=4.519\).
Indeed, the two initial perturbations \(\Delta\mathbf{x}_{17}^{A}\) and \(\Delta\mathbf{x}_{17}^{B}\) have the same perturbation energy, \(E_{17}(\Delta\mathbf{x}_{17}^{A})=E_{17}(\Delta\mathbf{x}_{17}^{B})=500\), but produce different outcomes because they are applied along different perturbation directions.
A larger \(\Delta_i\) therefore corresponds to a broader energy range over which perturbations with comparable energies can lead to different outcomes across directions, implying that robustness inferred from perturbations tested along only a limited set of directions may not represent the energy required to remain secured against perturbations from other directions.

Fig.~\ref{fig:stability_safety}c,d extend the comparison to all nodes across the seven power-grid models, with Fig.~\ref{fig:stability_safety}c comparing basin stability \(p_i^{\mathrm{BS}}\) and the relative secured-basin volume \(p_i^{\mathrm{S}}\).
Although the two measures show an overall positive association, the relationship remains widely dispersed, and the relatively low Pearson correlation \(r=0.457\) (\(R^2=0.209\)) indicates that they capture substantially different aspects of robustness.
In other words, perturbations classified as stable by basin stability based only on the synchronization outcome can be further distinguished according to whether their trajectories remain within the secured operating region.
Thus, a measure based only on the final synchronization outcome cannot fully characterize transient robustness, whereas the secured-basin classification reveals vulnerabilities that remain hidden when robustness is assessed solely by basin stability.

Fig.~\ref{fig:stability_safety}d further shows that the mean barrier height \(\overline{E}_i\) captures an aspect of robustness distinct from basin stability \(p_i^{\mathrm{BS}}\).
The two measures exhibit little linear association, with \(r=0.084\) (\(R^2=0.007\)), and nodes with similar basin stability can have substantially different characteristic energy scales.
One possible contributing factor is that, although both measures are evaluated within an analyst-specified perturbation region, basin stability quantifies the relative volume of perturbations that recover synchrony and is therefore more directly dependent on the chosen perturbation range, whereas \(\overline{E}_i\) characterizes the energy scale associated with the secured-basin boundary and is comparatively less sensitive to that range.
These results show that basin stability \(p_i^{\mathrm{BS}}\) and the mean barrier height \(\overline{E}_i\) provide complementary information on different aspects of robustness.

\subsection{Distinct structural and dynamical associations of energy-barrier measures}
\label{subsec:topology}

\begin{figure}
    \centering
    \includegraphics[width=0.9\textwidth]{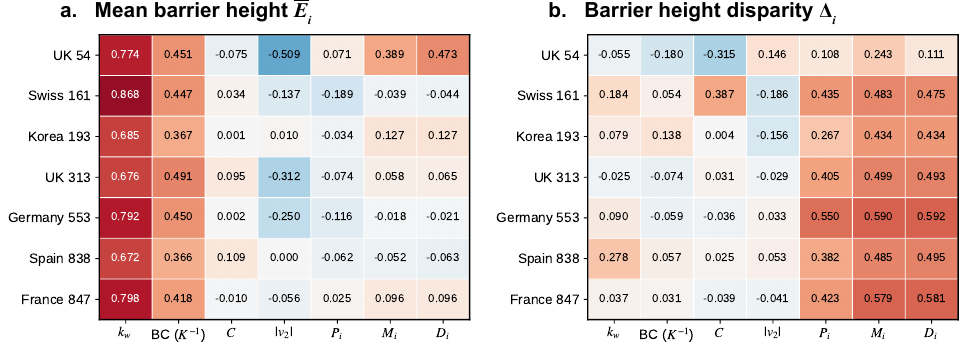}
    \caption{
    Node-wise Pearson correlations of the energy-barrier measures with structural measures and dynamical parameters across the seven power-grid networks.
    (a) Mean barrier height \(\overline{E}_i\).
    (b) Barrier height disparity \(\Delta_i\).
    The two measures exhibit distinct correlation patterns, with \(\overline{E}_i\) more consistently associated with \(k_w\) and \(BC(K^{-1})\), and \(\Delta_i\) with \(P_i\), \(M_i\), and \(D_i\).
    Each row corresponds to one power-grid network, and colors and numerical values indicate the Pearson correlation coefficients calculated within each network.
    }
    \label{fig:topology}
\end{figure}

We next examine what physical information is represented by the proposed node-level energy-barrier measures, the mean barrier height \(\overline{E}_i\) and barrier height disparity \(\Delta_i\), which capture aspects of robustness distinct from basin stability \(p_i^{\mathrm{BS}}\).
To evaluate their node-level associations with structural and dynamical properties, we use the same seven power-grid models analyzed in Sec.~\ref{subsec:high_stability}~\cite{lee2024reinforcement,tyloo2019key,pagnier2019inertia,song2025kpg}.
We consider four structural measures, the weighted degree \(k_w\), betweenness centrality \(BC(K^{-1})\), clustering coefficient \(C\), and absolute Fiedler vector component \(|v_2|\)  (see Sec.~\ref{subsec:methods_topology}), together with three dynamical parameters, the power injection \(P_i\), inertia \(M_i\), and damping coefficient \(D_i\), and evaluate their Pearson correlations with \(\overline{E}_i\) and \(\Delta_i\).
Fig.~\ref{fig:topology} compares these correlations across the seven power-grid models.


Fig.~\ref{fig:topology}a,b show that the two energy-barrier measures, \(\overline{E}_i\) and \(\Delta_i\), exhibit distinct association patterns.
The mean barrier height \(\overline{E}_i\) shows more consistent cross-node associations with \(k_w\) and \(BC(K^{-1})\) among the structural measures, whereas its correlations with the dynamical parameters \(P_i\), \(M_i\), and \(D_i\) are generally weak.
In contrast, the barrier height disparity \(\Delta_i\) shows little evidence of consistent cross-node associations with the considered node-level structural measures, but comparatively consistent cross-node associations with \(P_i\), \(M_i\), and \(D_i\).
These results indicate that \(\overline{E}_i\) is more closely associated with the structural properties of the node, whereas \(\Delta_i\) is more closely associated with its dynamical parameters.

The distinct correlation patterns of \(\overline{E}_i\) and \(\Delta_i\) can be interpreted in terms of the different physical information represented by the two measures.
For simplicity, consider the case in which the transition between secured and unsecured outcomes is primarily determined by the perturbed node itself.
The direct dependence of the potential-energy contribution on the coupling strengths \(K_{ij}\) is consistent with the positive association between \(\overline{E}_i\) and the weighted degree \(k_{w,i}=\sum_j K_{ij}\) (see Eq.~\ref{eq:potential_E} in Sec.~\ref{subsec:methods_energy}).
The association with \(BC(K^{-1})\) may additionally reflect network-mediated redistribution of perturbation energy, because nodes lying frequently on strongly coupled shortest paths may participate more strongly in energy transfer through the network.
In comparison, no consistent association is observed with the clustering coefficient \(C_i\) or the absolute Fiedler vector component \(|v_{2,i}|\), suggesting that local neighbor connectivity and global spectral position are less directly related to the overall energy-barrier scale.

By contrast, \(\Delta_i\) is defined from the ratio \(E_i^{\mathrm{U}}/E_i^{\mathrm{S}}\) through Eq.~\ref{eq:disparity}, so structural effects that shift the two energy thresholds in similar proportions can be partly cancelled in the ratio.
At the same time, the directional energy barrier \(E_i^{\mathrm{B}}(\hat{\mathbf d}_i)\) varies with the perturbation direction \(\hat{\mathbf d}_i\), reflecting differences in the energy required to reach the secured-basin boundary.
This variation can arise because the transient trajectory toward the boundary differs across perturbation directions, so the amounts of energy dissipated by damping and transferred to neighboring nodes can also differ.
Because the dynamical parameters \(P_i\), \(M_i\), and \(D_i\) affect these transient trajectories, their influence can remain reflected in \(\Delta_i\), which characterizes the directional heterogeneity of the energy barrier.

These interpretations describe the observed trends rather than establishing general causal relationships.
To examine whether controlled changes in nodal structural connectivity and dynamical parameters induce corresponding changes in these metrics, we performed a sensitivity analysis for node \(17\) of the 54-node British power grid by independently varying the coupling strengths of its adjacent links and its nodal dynamical parameters by \(\pm 50\%\).
As summarized in Table~\ref{tab:sensitivity_17}, basin stability \(p_{17}^{\mathrm{BS}}\) remained unchanged under all conditions, while the relative secured-basin volume \(p_{17}^{\mathrm{S}}\) responded much more strongly to changes in the nodal dynamical parameters than to coupling variations.
By contrast, the mean barrier height \(\overline{E}_{17}\) and barrier height disparity \(\Delta_{17}\) showed larger responses to coupling variations, although only the response of \(\overline{E}_{17}\) was consistent with the correlation pattern observed in Fig.~\ref{fig:topology}.
These results indicate that the associations observed across nodes do not necessarily predict how an individual node responds to controlled parameter changes.

\begin{table}
\centering
\begin{tabular}{lcccc}
\hline
Condition & \(p_{17}^{\mathrm{BS}}\) & \(p_{17}^{\mathrm{S}}\) & \(\overline{E}_{17}\) & \(\Delta_{17}\) \\
\hline
Reference case & \(1.000\;(-)\) & \(0.308\;(-)\) & \(239.880\;(-)\) & \(4.519\;(-)\) \\
Coupling strength \(+50\%\) (\(k_{w,17}\)) & \(1.000\;(0.0\%)\) & \(0.307\;(-0.5\%)\) & \(294.346\;\mathbf{(+22.7\%)}\) & \(5.460\;\mathbf{(+20.8\%)}\) \\
Coupling strength \(-50\%\) (\(k_{w,17}\)) & \(1.000\;(0.0\%)\) & \(0.312\;(+1.2\%)\) & \(175.419\;\mathbf{(-26.9\%)}\) & \(3.306\;\mathbf{(-26.9\%)}\) \\
Dynamical parameters \(+50\%\) (\(P_{17}\), \(M_{17}\), \(D_{17}\)) & \(1.000\;(0.0\%)\) & \(0.223\;\mathbf{(-27.7\%)}\) & \(220.502\;(-8.1\%)\) & \(4.337\;(-4.0\%)\) \\
Dynamical parameters \(-50\%\) (\(P_{17}\), \(M_{17}\), \(D_{17}\)) & \(1.000\;(0.0\%)\) & \(0.619\;\mathbf{(+101.1\%)}\) & \(291.310\;(+21.4\%)\) & \(4.289\;(-5.1\%)\) \\
\hline
\end{tabular}
\caption{ 
Sensitivity of basin stability \(p_{17}^{\mathrm{BS}}\), relative secured-basin volume \(p_{17}^{\mathrm{S}}\), mean barrier height \(\overline{E}_{17}\), and barrier height disparity \(\Delta_{17}\) of node \(17\) in the 54-node British power grid to variations in coupling strength and nodal dynamical parameters.
Values in parentheses indicate relative changes from the reference case.
Within each metric, the two largest absolute percentage changes across the parameter-variation conditions are highlighted in boldface.
}
\label{tab:sensitivity_17}
\end{table}

\subsection{Transferability of energy thresholds between perturbation representations}
\label{subsec:power}

In the preceding analyses, perturbations were represented as instantaneous changes \(\Delta\theta_i\) and \(\Delta\omega_i\) in the phase angle and angular velocity of a target node \(i\), respectively.
Although this initial-condition framework is useful for characterizing nonlinear stability, disturbances in practical power-grid operation more commonly arise as finite-duration power imbalances, including faults, load changes, and generation fluctuations~\cite{kundur2007power,machowski2020power,nguyen2015voltage}.
Nevertheless, the energy thresholds obtained in the state-perturbation space remain physically meaningful.
Although state perturbations and finite-duration power disturbances are represented in different coordinates and enter the system in different forms, both inject energy into the system through the perturbed oscillator.
Hereafter, we distinguish between state perturbations, which directly modify the initial phase angle and angular velocity, and power disturbances, which act on the system as finite-duration power imbalances in a similar sense of the previous study~\cite{lee2017uniformity}.
This common energy representation allows us to examine whether the secured and unsecured energy thresholds, \(E_i^{\mathrm{S}}\) and \(E_i^{\mathrm{U}}\), obtained in the \((\Delta\theta_i,\Delta\omega_i)\) state-perturbation plane retain their meaning as thresholds in the \((\Delta P_i,\Delta t_i)\) power-disturbance plane.

\begin{figure}
    \centering
    \includegraphics[width=0.8\textwidth]{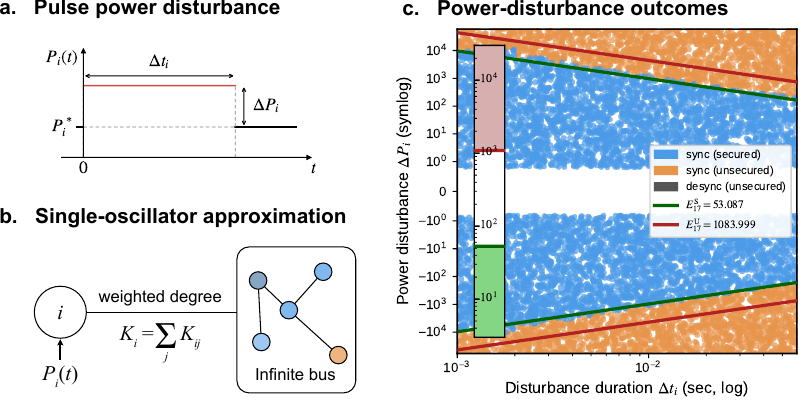}
    \caption{
    Finite-duration power disturbances and their outcomes for node \(17\) of the 54-node British power grid.
    (a) Pulse power disturbance with magnitude \(\Delta P_i\) and duration \(\Delta t_i\).
    (b) Equivalent single-oscillator representation of node \(i\) coupled to an infinite bus with \(K_i=\sum_j K_{ij}\).
    (c) Power-disturbance outcomes in the \((\Delta P_i,\Delta t_i)\) plane.
    Blue and orange points denote secured and unsecured synchronized outcomes, respectively; unsecured desynchronized outcomes are absent for this node.
    The green and red curves are the constant-energy contours obtained by mapping the state-derived thresholds \(E_{17}^{\mathrm{S}}=53.087\) and \(E_{17}^{\mathrm{U}}=1083.999\), respectively, onto the \((\Delta P_i,\Delta t_i)\) plane.
    The vertical bar indicates the secured, mixed-outcome, and unsecured energy ranges defined by these same state-derived thresholds.
    }
    \label{fig:fault}
\end{figure}

We consider pulse power disturbances characterized by their magnitude \(\Delta P_i\) and duration \(\Delta t_i\), as illustrated in Fig.~\ref{fig:fault}a.
Using the short-duration energy mapping derived in Sec.~\ref{subsec:methods_power_disturbance} under the equivalent infinite-bus approximation illustrated in Fig.~\ref{fig:fault}b, the state-derived thresholds \(E_i^{\mathrm{S}}\) and \(E_i^{\mathrm{U}}\) can be mapped onto the \((\Delta P_i,\Delta t_i)\) plane as constant-energy contours.
Fig.~\ref{fig:fault}c shows the sampled power-disturbance outcomes for node \(17\) of the 54-node British power grid together with the boundary contours obtained by substituting \(E_i=E_i^{\mathrm{S}}\) and \(E_i=E_i^{\mathrm{U}}\) into the power-disturbance mapping.
Secured outcomes are concentrated predominantly inside the \(E_{17}^{\mathrm{S}}\) contour, whereas unsecured outcomes dominate outside the \(E_{17}^{\mathrm{U}}\) contour, with both outcome types observed between the two contours.
For this representative node, the state-derived thresholds therefore provide a close empirical correspondence with the sampled separation of secured, mixed-outcome, and unsecured regions in the power-disturbance plane.

To examine whether this correspondence extends beyond the representative node, we evaluated the threshold violation rate
\begin{equation}
\epsilon_i
=
\frac{N_i^{\mathrm{miss}}}
{N_i^{\mathrm{match}}+N_i^{\mathrm{miss}}},
\label{eq:threshold_violation_result}
\end{equation}
for all nodes across the seven power-grid models.
Here, \(N_i^{\mathrm{match}}\) and \(N_i^{\mathrm{miss}}\) denote the numbers of sampled power-disturbance outcomes that agree and disagree with the state-derived energy-threshold classification, respectively (see Sec.~\ref{subsec:methods_power_disturbance} for the detailed classification).
Thus, \(\epsilon_i\) is an empirical consistency measure defined with respect to the prescribed power-disturbance sampling ensemble, rather than a probability weighted by the occurrence likelihood of disturbances.
It quantifies the fraction of sampled outcomes outside the mixed-outcome interval \(\mathcal{M}_i\) that are inconsistent with the corresponding state-derived energy thresholds.

The resulting statistics of \(\epsilon_i\) are summarized in Table~\ref{tab:power_violation}.
Across the seven power-grid models, the mean violation rates range from \(1.177\%\) to \(2.118\%\), indicating that the state-derived thresholds remain broadly consistent with the sampled short-duration power disturbances across the tested networks.
The observed violations may partly arise from the approximations used to map the state-derived energy thresholds into the power-disturbance plane, including the equivalent infinite-bus and short-duration assumptions illustrated in Fig.~\ref{fig:fault}b.

\begin{table}[t]
\centering
\begin{tabular}{lccccccc}
\hline
Statistic & UK 54 & Swiss 161 & Korea 193 & UK 313 & Germany 553 & Spain 838 & France 847 \\
\hline
Mean & 1.177\% & 2.118\% & 1.533\% & 1.745\% & 1.955\% & 1.923\% & 1.753\% \\
Standard deviation & 0.838\% & 1.838\% & 1.164\% & 1.142\% & 1.505\% & 1.670\% & 1.325\% \\\hline
\end{tabular}
\caption{
Mean and standard deviation of the threshold violation rate \(\epsilon_i\) across nodes in each of the seven power-grid models.
}
\label{tab:power_violation}
\end{table}

The power-disturbance representation also provides a direct interpretation in terms of practical fault-clearing time in power grids.
For node \(17\) of the 54-node British power grid shown in Fig.~\ref{fig:fault}c, considering a fixed disturbance magnitude \(|\Delta P_{17}|\) corresponds to taking a horizontal cross-section of the \((\Delta P_{17},\Delta t_{17})\) plane.
The intersections of this cross-section with the \(E_{17}^{\mathrm{S}}\) and \(E_{17}^{\mathrm{U}}\) contours represent characteristic disturbance durations separating the secured-only, mixed-outcome, and unsecured-only regions.
These intersections therefore indicate, for a given power-disturbance magnitude, the mixed-outcome duration ranges over which secured operation is maintained or lost.
Furthermore, the state-derived energy thresholds provide power-dependent duration boundaries without requiring the stability threshold to be identified independently for each disturbance magnitude.

\subsection{Boundary-focused search for efficient energy-threshold estimation}
\label{subsec:boundary_focused_search}

In addition to demonstrating the practical transferability of the energy thresholds between perturbation representations in the preceding analysis, we next examine whether the boundary-focused search introduced in Sec.~\ref{subsubsec:efficient_threshold_identification} is effective for estimating the energy thresholds \(E_i^{\mathrm{S}}\) and \(E_i^{\mathrm{U}}\).
Specifically, we compare the boundary-focused search with uniform random sampling in terms of how rapidly the secured and unsecured energy-threshold estimates converge as the number of evaluated perturbations increases.
Since the exact threshold energies \(E_i^{\mathrm{S}}\) and \(E_i^{\mathrm{U}}\) are not directly available, we define empirical reference thresholds for each node using the final estimates obtained after 1,000 evaluated perturbations with the two sampling strategies.
The smaller of the two final estimates is taken as \(E_{i,\mathrm{ref}}^{\mathrm{S}}\), whereas the larger is used for \(E_{i,\mathrm{ref}}^{\mathrm{U}}\).
At each iteration, both sampling strategies are evaluated against the same node-wise reference thresholds, and their deviations are quantified using the normalized mean absolute error (NMAE).
After \(k\) evaluated perturbations, the NMAEs are defined as
\begin{equation}
\mathrm{NMAE}^{\mathrm{S}}(k)
=
\frac{1}{N}
\sum_{i=1}^{N}
\frac{
\left|
E_{i,k}^{\mathrm{S}}-E_{i,\mathrm{ref}}^{\mathrm{S}}
\right|
}{
E_{i,\mathrm{ref}}^{\mathrm{S}}
},
\end{equation}
and
\begin{equation}
\mathrm{NMAE}^{\mathrm{U}}(k)
=
\frac{1}{N}
\sum_{i=1}^{N}
\frac{
\left|
E_{i,k}^{\mathrm{U}}-E_{i,\mathrm{ref}}^{\mathrm{U}}
\right|
}{
E_{i,\mathrm{ref}}^{\mathrm{U}}
},
\end{equation}
where \(E_{i,k}^{\mathrm{S}}\) and \(E_{i,k}^{\mathrm{U}}\) denote the current estimates after \(k\) evaluated perturbations, and \(E_{i,\mathrm{ref}}^{\mathrm{S}}\) and \(E_{i,\mathrm{ref}}^{\mathrm{U}}\) denote the corresponding empirical reference thresholds.

\begin{figure}
    \centering
    \includegraphics[width=1.0\textwidth]{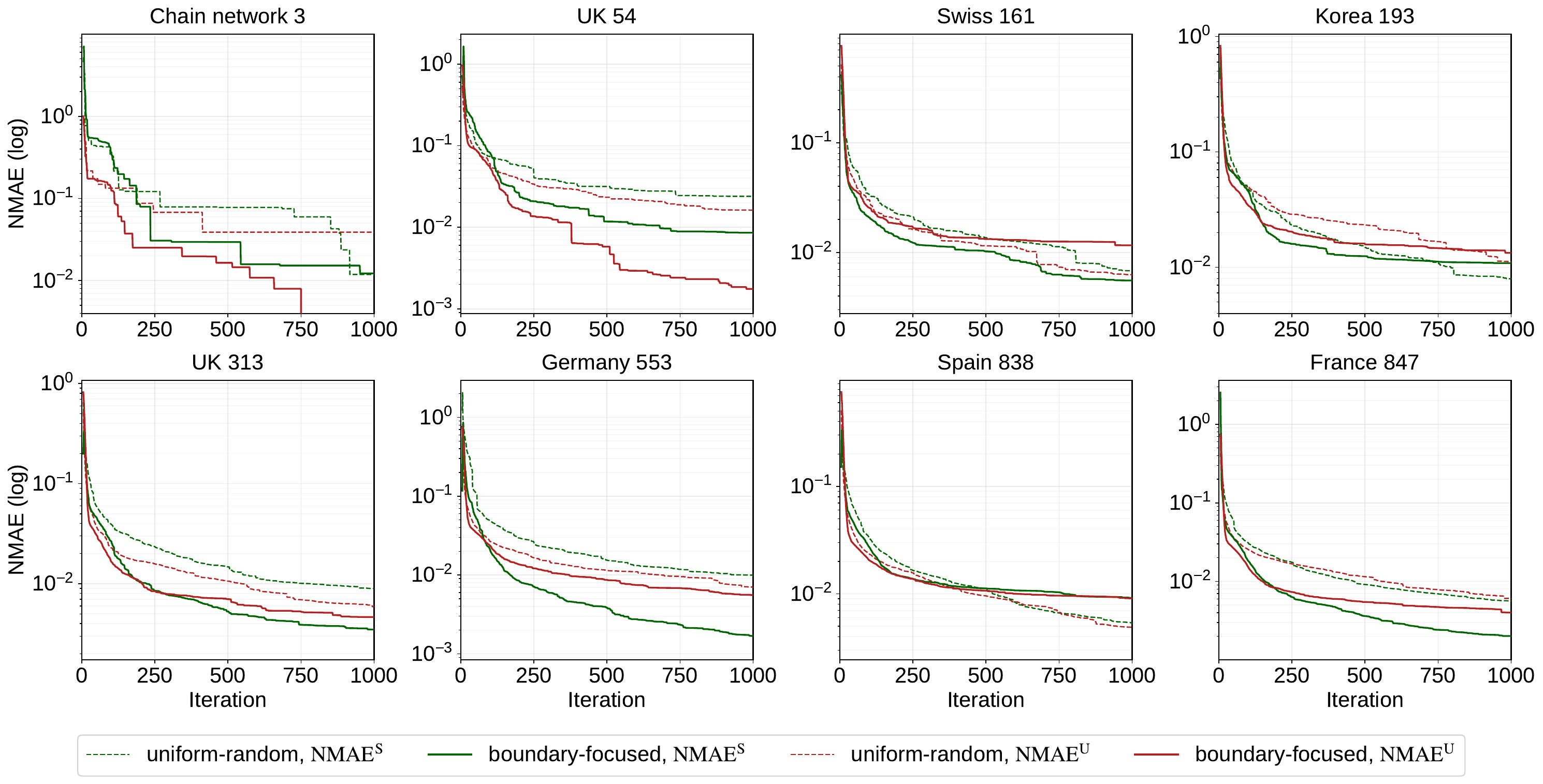}
\caption{
Convergence of secured and unsecured energy-threshold estimates under uniform random sampling and the boundary-focused search.
The panels correspond to the 3-node chain network and seven power-grid models. Each panel shows the normalized mean absolute error (NMAE) of \(E_{i,k}^{\mathrm{S}}\) and \(E_{i,k}^{\mathrm{U}}\) with respect to the final estimates after 1,000 evaluations, as the number of evaluated perturbations increases.
The smaller final estimate from the two sampling strategies is used for \(E_{i,\mathrm{ref}}^{\mathrm{S}}\), and the larger for \(E_{i,\mathrm{ref}}^{\mathrm{U}}\).
Green and red curves denote the secured and unsecured thresholds, respectively, while dashed and solid curves indicate uniform random sampling and the boundary-focused search.
}
    \label{fig:fast}
\end{figure}

Fig.~\ref{fig:fast} compares the two sampling strategies using the 3-node chain network introduced in Fig.~\ref{fig:basin_safety} and the seven power-grid models analyzed throughout this study~\cite{lee2024reinforcement,tyloo2019key,pagnier2019inertia,song2025kpg}.
Across all cases in Fig.~\ref{fig:fast}, the boundary-focused search shows comparable or lower errors than uniform random sampling during the early stage of sampling, and this advantage consistently persists down to the \(2\%\) error level.
Table~\ref{tab:fig7_convergence} further shows that the boundary-focused search generally requires fewer iterations across the tested error levels, including the \(1\%\) level.
At the \(1\%\) error level, however, the extent of this error reduction varies across networks and between the two thresholds.
Uniform random sampling becomes more efficient for the Swiss, Korea, and Spanish grids, indicating that the advantage of the boundary-focused search is not uniform across all networks as higher accuracy is required.

\begin{table}[t]
\centering
\begin{tabular}{l cccc cccc c}
\hline
 & \multicolumn{4}{c}{Uniform random sampling} & \multicolumn{4}{c}{Boundary-focused search} & \\
\cmidrule(lr){2-5} \cmidrule(lr){6-9}
Network & 5\% & 3\% & 2\% & 1\% & 5\% & 3\% & 2\% & 1\% & Crossing \\
\hline
Chain network 3 & 852 & --- & --- & --- & \textbf{235} & \textbf{307} & \textbf{544} & --- & --- \\
UK 54 & 250 & 517 & --- & --- & \textbf{120} & \textbf{184} & \textbf{284} & \textbf{681} & --- \\
Swiss 161 & 66 & 131 & 254 & \textbf{806} & \textbf{28} & \textbf{79} & \textbf{156} & --- & 541 (1.3\%) \\
Korea 193 & 98 & 224 & 649 & --- & \textbf{88} & \textbf{130} & \textbf{256} & --- & 790 (1.4\%) \\
UK 313 & 70 & 161 & 311 & 789 & \textbf{41} & \textbf{88} & \textbf{112} & \textbf{223} & --- \\
Germany 553 & 97 & 193 & 357 & 972 & \textbf{58} & \textbf{78} & \textbf{116} & \textbf{372} & --- \\
Spain 838 & 64 & 114 & 192 & \textbf{523} & \textbf{40} & \textbf{91} & \textbf{135} & 768 & 494 (1.1\%) \\
France 847 & 52 & 103 & 196 & 561 & \textbf{25} & \textbf{67} & \textbf{94} & \textbf{158} & --- \\
\hline
Weighted average & 73.9 & 146.9 & 285.2 & 701.4 & \textbf{43.3} & \textbf{85.2} & \textbf{129.5} & \textbf{488.4} &  \\
\hline
\end{tabular}
\caption{
Number of iterations required for the NMAEs of both the secured and unsecured energy-threshold estimates, \(E_{i,k}^{\mathrm{S}}\) and \(E_{i,k}^{\mathrm{U}}\), to fall below the specified error levels under uniform random sampling and the boundary-focused search, with a common budget of 1,000 iterations per method.
The smaller final estimate from the two sampling strategies is used for \(E_{i,\mathrm{ref}}^{\mathrm{S}}\), and the larger for \(E_{i,\mathrm{ref}}^{\mathrm{U}}\).
Bold values indicate the method requiring fewer iterations, and ``---'' indicates that the specified error level was not reached within 1,000 iterations.
The Crossing column gives the last iteration after which uniform random sampling remains more accurate than the boundary-focused search, with the boundary-focused NMAE at that iteration shown in parentheses.
The bottom row reports node-count-weighted averages across all networks, with ``---'' entries treated as 1,000 iterations.
}
\label{tab:fig7_convergence}
\end{table}

The advantage of the boundary-focused search arises from selecting perturbations near the current energy thresholds rather than sampling arbitrary state perturbations across the entire perturbation domain.
By concentrating simulations in energy ranges that are likely to update \(E_{i,k}^{\mathrm{S}}\) and \(E_{i,k}^{\mathrm{S}}\), the thresholds can be estimated efficiently without uniformly exploring the full perturbation domain.
However, the initial perturbation energy \(E_i(\Delta\mathbf{x}_i)\) defined in Eq.~\ref{eq:perturbation_energy} does not capture the relatively small differences arising from energy dissipated by damping or transferred to other nodes during the transient.
These omitted transient effects may become more relevant at very low error levels, which can reduce or even reverse the advantage of the boundary-focused search in some networks as higher accuracy is required.
Overall, the boundary-focused search tends to substantially reduce the simulation budget required to obtain threshold estimates at a practically useful level of accuracy.

\section{Discussion}
\label{sec:discussion}

In this study, we defined an admissible operating region for an inertially coupled oscillator model by imposing a phase-angle condition that preserves the physical interpretability of its trajectories as responses of the actual power grid, and defined the secured basin of each node as the region in perturbation space formed by state perturbations whose trajectories remain within this region throughout the transient.
The secured basin of each node provides an identifiable boundary in perturbation space.
Expressing the perturbations along this boundary on a common physical energy scale allows the perturbation strength required to leave the secured basin to serve as a quantitative measure of node-level synchronization stability.
We interpret these boundary energies as energy barriers and summarize their node-wise distribution using two measures, the mean barrier height \(\overline{E}_i\), representing the characteristic energy scale, and the barrier height disparity \(\Delta_i\), representing its heterogeneity.
Because these measures are determined by the extrema of the energy-barrier distribution, we further developed a practical strategy for estimating them with reduced computational cost.


The secured-basin framework provides a physically meaningful basis for assessing transient robustness beyond the final synchronization state. By distinguishing the region in which the modeled response remains physically interpretable as the behavior of the original grid, it identifies transient vulnerabilities that would be missed if stability were assessed solely from whether trajectories eventually resynchronize. The resulting energy-barrier characterization quantifies both the amount of perturbation a node can withstand and the extent to which this tolerance depends on perturbation direction, while reducing sensitivity to the chosen perturbation range when the relevant secured-basin boundary is sufficiently contained. Moreover, the mean barrier height \(\overline{E}_i\) and barrier height disparity \(\Delta_i\) exhibit distinct node-wise association patterns with structural connectivity and dynamical parameters, providing complementary information on the factors related to node-level transient robustness.

The framework also offers practical advantages in both interpretation and computation. Energy thresholds identified from state-space perturbations remain effective when disturbances are instead represented as finite-duration power imbalances, indicating that the energy-based characterization provides a common physical basis across different perturbation representations. Their mapping to power--duration coordinates further connects these thresholds to considerations related to fault-clearing time in power grids. Finally, because the relevant barrier measures are determined by the portions of the secured-basin boundary that set the energy thresholds rather than by the basin volume as a whole, they can be estimated efficiently by concentrating simulations near these regions instead of sampling the entire perturbation space uniformly.

Despite these advantages, the present study is limited to single-node perturbations, which are useful for isolating node-level vulnerabilities but do not capture all disturbances relevant to power-grid operation.
Faults and outages can also alter link states, network topology, and coupling strengths, and therefore cannot always be represented by perturbations to nodal states or power injections alone.
Extending the secured-basin framework to combined node and link perturbations that induce structural changes in the network remains an important direction for future work.

Beyond power-grid applications, the framework can be extended to other synchronized dynamical systems in which transient trajectories are required to remain within admissible regions of state space.
Although the relevant state constraints and disturbance mechanisms may differ across systems, the same energy-barrier characterization can quantify both the perturbation scale required to exit the admissible region and the extent to which this scale varies with perturbation direction.
This offers a general framework for linking transient stability boundaries to system-specific disturbances through a common energy-based description.

\section{Methods}
\label{sec:methods}

\subsection{Power-grid dynamics}
\label{subsec:methods_model}

The second-order Kuramoto model is widely used to study collective dynamics arising from inertia, damping, and nonlinear coupling~\cite{tanaka1997first,tanaka1997self,olmi2014hysteretic,gao2018self,gao2021synchronized,kim2025cluster,kim2026paths,kim2026synchronization,kim2026predicting,yi2026phase,kim2026decentralization, kim2026frequency}.
We model the power grid within this framework as a network of synchronous machines coupled through transmission lines.
Let \(\theta_i\) and \(\omega_i=\dot{\theta}_i\) denote the phase angle and angular velocity of node \(i\), respectively.
The dynamics are given by
\[
\dot{\theta}_i=\omega_i~,
\]
\[
M_i\dot{\omega}_i
=
P_i
-
D_i\omega_i
+
\sum_j K_{ij}\sin(\theta_j-\theta_i).
\]
Here, \(P_i\) is the net power injection at node \(i\), \(M_i\) is the inertia, \(D_i\) is the damping coefficient, and \(K_{ij}\) is the coupling strength between nodes \(i\) and \(j\).
All power grids are represented as undirected weighted networks, with \(K_{ij}=0\) for node pairs that are not directly connected.

The net power injection at each node is defined as the difference between generation and load.
The coupling strength of each transmission line is taken to be proportional to the inverse of its reactance.
For parallel lines connecting the same pair of nodes, the corresponding coupling strengths are summed, while inactive lines are excluded.
When inertia or damping data are unavailable, \(M_i\) and \(D_i\) are estimated from nodal power injection using regression models~\cite{park2025optimal}.
These preprocessing steps place all power-grid datasets within a common swing-equation representation.

The synchronized operating point of each power grid is denoted by
\(
\mathbf{x}^{\ast}
=
(\boldsymbol{\theta}^{\ast},\boldsymbol{0})
\)
and is obtained by numerically solving the steady-state swing equations for the specified power injections and coupling strengths.
All state perturbations and power disturbances are defined relative to this operating point.
All trajectories are integrated using a fourth-order Runge--Kutta scheme with a time step of \(10^{-3}\,\mathrm{s}\).
Each simulation comprises \(500{,}000\) time steps, corresponding to a total duration of \(500\,\mathrm{s}\).

\subsection{Outcome classification under state perturbations}
\label{subsec:methods_state_perturbation}

To assess node-level vulnerability, we apply single-node state perturbations.
For a target node \(i\), only its initial phase-angle and angular-velocity are perturbed, while all other nodes remain at the synchronized operating point.
The initial conditions are given by
\[
\theta_i(0)
=
\theta_i^{\ast}
+
\Delta\theta_i,
\qquad
\omega_i(0)
=
\Delta\omega_i,
\]
\[
\theta_j(0)
=
\theta_j^{\ast},
\qquad
\omega_j(0)
=
0
\quad
(j\neq i).
\]
For the analyses based on uniform random sampling, the perturbations are sampled uniformly in the geometric \((\Delta\theta_i,\Delta\omega_i)\) coordinates over \(\Delta\theta_i\in[-\pi,\pi]\) and \(\Delta\omega_i\in[-20,20]\).
The perturbations are sampled from the perturbation set \(A_i\), which is used to estimate the basin stability \(p_i^{\mathrm{BS}}\) and relative secured-basin volume \(p_i^{\mathrm{S}}\).

For each sampled perturbation, the swing equations are integrated numerically and the resulting trajectory is classified according to both its final synchronization state and whether it remains within the secured operating region \(X_{\mathrm{secured}}\) defined in Eq.~\ref{eq:X_secured}.
A trajectory is classified as synchronized if \(|\omega_j(T)|<10^{-3}\) for every node \(j\) at the end of the observation horizon.
Perturbations whose trajectories satisfy the secured operating criterion are assigned to \(\mathcal{S}_i\), whereas all remaining perturbations are assigned to \(\mathcal{U}_i\).
Because the secured operating criterion uses the symmetric modulo operation defined in Eq.~\ref{eq:X_secured}, crossings of the phase-angle boundary at \(\Delta\theta_{\mathrm{crit}}=\pi\) are detected numerically from transitions across its branch cut between consecutive integration steps.
The basin stability \(p_i^{\mathrm{BS}}\) and relative secured-basin volume \(p_i^{\mathrm{S}}\) are then estimated from the corresponding fractions of uniformly sampled perturbations in \(A_i\).

\subsection{Energy mapping of state perturbations}
\label{subsec:methods_energy}

The perturbation energy $E_i(\Delta\mathbf{x}_i)$ defined in Eq.~\ref{eq:perturbation_energy} is evaluated from the change in kinetic and potential energy produced by the state perturbation applied to node $i$.
Because the synchronized operating point satisfies $\omega_i^{\ast}=0$, the kinetic-energy change associated with the angular-velocity perturbation is
\begin{equation}
\Delta T_i(\Delta\omega_i)
=
\frac{1}{2}M_i(\Delta\omega_i)^2.
\end{equation}
The potential-energy change associated with the phase-angle perturbation is
\begin{equation}
\Delta V_i(\Delta\theta_i)
=
-
\sum_j K_{ij}
\left[
\cos\left(
\theta_i^{\ast}
+
\Delta\theta_i
-
\theta_j^{\ast}
\right)
-
\cos\left(
\theta_i^{\ast}
-
\theta_j^{\ast}
\right)
\right]
-
P_i\Delta\theta_i.
\label{eq:potential_E}
\end{equation}
This expression gives the change in the conservative potential associated with displacing node $i$ from its synchronized operating point while the initial states of all other nodes remain fixed at their operating-point values.

The resulting perturbation energy is therefore evaluated as
\[
E_i(\Delta\mathbf{x}_i)
=
\Delta T_i(\Delta\omega_i)
+
\Delta V_i(\Delta\theta_i),
\]
consistent with Eq.~\ref{eq:perturbation_energy}.
Damping affects the subsequent trajectory and therefore the classification into $\mathcal{S}_i$ or $\mathcal{U}_i$, but it is not included in $E_i(\Delta\mathbf{x}_i)$ because it represents a nonconservative contribution.
Accordingly, $E_i(\Delta\mathbf{x}_i)$ is not a conserved quantity under the full damped dynamics, but is used as a common energy scale for comparing state perturbations and identifying the energy thresholds defined in Eqs.~\ref{eq:secured_energy} and \ref{eq:unsecured_energy}.

To determine the phase-angle range used for the energy-barrier analysis, we estimate a local unstable equilibrium for each perturbed node while fixing the neighboring nodes at their synchronized phase angles.
The coupling terms are represented by the local mean field
\begin{equation}
R_i e^{\mathrm{i}\psi_i}
=
\sum_j K_{ij} e^{\mathrm{i}\theta_j^\ast},
\end{equation}
so that the equilibrium condition of node \(i\) becomes
\begin{equation}
P_i
=
R_i\sin(\theta_i-\psi_i).
\label{eq:local_uep}
\end{equation}
Among the two solutions of Eq.~\ref{eq:local_uep}, one corresponds to the reference stable equilibrium and the other is taken as the local unstable equilibrium.
Phase perturbations extending beyond this unstable-equilibrium angle are excluded from the energy-barrier analysis.

\subsection{Boundary-focused search}
\label{subsec:methods_search}

The boundary-focused search described in Sec.~\ref{subsubsec:efficient_threshold_identification} is used to estimate the two energy thresholds without uniformly sampling the entire perturbation plane.
For a target node \(i\), the search is initialized using the unperturbed state and two corner perturbations located on the side of the principal phase interval that remains after applying the unstable-equilibrium restriction.
These three deterministic perturbations are
\begin{equation}
\Delta\mathbf{x}_i=(0,0),
\qquad
\Delta\mathbf{x}_i=(\Delta\theta_i^{\mathrm{far}},\pm\omega_{\max}),
\end{equation}
where \(\Delta\theta_i^{\mathrm{far}}=\pi\) when the excluded unstable-equilibrium region lies on the negative side and \(\Delta\theta_i^{\mathrm{far}}=-\pi\) when it lies on the positive side.
After the three initial perturbations have been evaluated, the resulting temporary thresholds are denoted by \(E_{i,3}^{\mathrm{S}}\) and \(E_{i,3}^{\mathrm{U}}\).
More generally, after \(k\) perturbations have been evaluated, the temporary thresholds are denoted by \(E_{i,k}^{\mathrm{S}}\) and \(E_{i,k}^{\mathrm{U}}\), and the interval between them by \(\mathcal{M}_{i,k}\).

Before the ordering \(E_{i,k}^{\mathrm{S}}<E_{i,k}^{\mathrm{U}}\) is established, the temporary thresholds satisfy \(E_{i,k}^{\mathrm{U}}<E_{i,k}^{\mathrm{S}}\).
An expansion factor is defined as
\begin{equation}
r_{i,k}
=
\sqrt{
\frac{E_{i,k}^{\mathrm{S}}}
{E_{i,k}^{\mathrm{U}}}
},
\end{equation}
and target energies are sampled from
\begin{equation}
E_i^{\mathrm{target}}
\in
\left[
\frac{E_{i,k}^{\mathrm{U}}}{r_{i,k}},
E_{i,k}^{\mathrm{S}} \times r_{i,k}
\right].
\end{equation}

Once \(E_{i,k}^{\mathrm{S}}<E_{i,k}^{\mathrm{U}}\) is established, the interval
\(\mathcal{M}_{i,k}=[E_{i,k}^{\mathrm{S}},E_{i,k}^{\mathrm{U}}]\)
cannot directly update either threshold.
At this stage,
\begin{equation}
\Delta_{i,k}
=
\sqrt{
\frac{E_{i,k}^{\mathrm{U}}}
{E_{i,k}^{\mathrm{S}}}
},
\end{equation}
and the search is restricted to the two outer energy intervals
\begin{equation}
E_i^{\mathrm{target}}
\in
\left[
\frac{E_{i,k}^{\mathrm{S}}}{\Delta_{i,k}},
E_{i,k}^{\mathrm{S}}
\right]
\cup
\left[
E_{i,k}^{\mathrm{U}},
E_{i,k}^{\mathrm{U}} \times \Delta_{i,k}
\right].
\end{equation}

Within the current target-energy domain, \(E_i^{\mathrm{target}}\) is sampled log-uniformly.
The phase-angle perturbation \(\Delta\theta_i\) is then sampled uniformly from the portion of the principal interval \((-\pi,\pi]\) bounded by the local unstable equilibrium on one side and by \(\pm\pi\) on the other.
For the selected pair \((E_i^{\mathrm{target}},\Delta\theta_i)\), the corresponding angular-velocity perturbation is obtained by inverting the perturbation-energy relation:
\[
\Delta\omega_i
=
\pm
\sqrt{
\frac{2}{M_i}
\left[
E_i^{\mathrm{target}}
-
\Delta V_i(\Delta\theta_i)
\right]
}.
\]
The sign of \(\Delta\omega_i\) is selected with equal probability.
If \(E_i^{\mathrm{target}}<\Delta V_i(\Delta\theta_i)\), no real-valued angular-velocity perturbation exists, and both \(E_i^{\mathrm{target}}\) and \(\Delta\theta_i\) are resampled from their respective distributions.
The resulting perturbation \(\Delta\mathbf{x}_i=(\Delta\theta_i,\Delta\omega_i)\) therefore satisfies \(E_i(\Delta\mathbf{x}_i)=E_i^{\mathrm{target}}\).

Each sampled perturbation is simulated and classified into either \(\mathcal{S}_i\) or \(\mathcal{U}_i\).
If an unsecured perturbation satisfies \(E_i(\Delta\mathbf{x}_i)<E_{i,k}^{\mathrm{S}}\), the temporary secured threshold is updated to this lower value.
If a secured perturbation satisfies \(E_i(\Delta\mathbf{x}_i)>E_{i,k}^{\mathrm{U}}\), the temporary unsecured threshold is updated to this higher value.
The corresponding \(\mathcal{M}_{i,k}\) and \(\Delta_{i,k}\) are recalculated after each update.

The procedure is repeated until a prescribed total number of perturbations \(K\) has been evaluated.
The resulting \(E_{i,K}^{\mathrm{S}}\), \(E_{i,K}^{\mathrm{U}}\), \(\Delta_{i,K}\), and \(\mathcal{M}_{i,K}\) are taken as the finite-sample estimates of \(E_i^{\mathrm{S}}\), \(E_i^{\mathrm{U}}\), \(\Delta_i\), and \(\mathcal{M}_i\), respectively.

\subsection{Node-level structural properties}
\label{subsec:methods_topology}

For each node \(i\), we first calculate the weighted degree and betweenness centrality.
The weighted degree is defined as
\begin{equation}
k_{w,i}
=
\sum_j K_{ij},
\end{equation}
where \(K_{ij}\) denotes the coupling strength between nodes \(i\) and \(j\).
The betweenness centrality is calculated as
\begin{equation}
BC_i
=
\sum_{\substack{s\neq i\neq t\\s<t}}
\frac{\sigma_{st}(i)}{\sigma_{st}},
\end{equation}
where \(\sigma_{st}\) is the number of shortest paths between nodes \(s\) and \(t\), and \(\sigma_{st}(i)\) is the number of those paths passing through node \(i\).
For this calculation, the edge distance is defined as \(d_{ij}=K_{ij}^{-1}\), so that stronger coupling corresponds to a shorter network distance.

We additionally characterize the local and spectral structure around each node using the clustering coefficient and the Fiedler vector.
The local clustering coefficient is defined as
\begin{equation}
C_i
=
\frac{e_i}{\binom{k_i}{2}},
\end{equation}
where \(k_i\) is the degree of node \(i\), \(e_i\) is the number of edges among its neighboring nodes, and \(\binom{k_i}{2}\) is the maximum possible number of such edges.
For nodes with \(k_i<2\), we set \(C_i=0\).
To characterize the spectral position of each node, we construct the weighted graph Laplacian
\begin{equation}
\mathbf{L}
=
\mathbf{D}_K-\mathbf{K},
\end{equation}
where \(\mathbf{K}\) is the weighted adjacency matrix and \(\mathbf{D}_K\) is the diagonal matrix of weighted degrees.
The Fiedler vector \(\mathbf{v}_2\) is defined by
\begin{equation}
\mathbf{L}\mathbf{v}_2
=
\lambda_2\mathbf{v}_2,
\end{equation}
where \(\lambda_2\) is the second-smallest eigenvalue of \(\mathbf{L}\).
For node \(i\), we use the absolute value of its corresponding Fiedler-vector component, \(|v_{2,i}|\).

\subsection{Power-disturbance model and energy mapping}
\label{subsec:methods_power_disturbance}

In addition to state perturbations, we consider finite-duration power disturbances applied to a target node \(i\) (see Fig.~\ref{fig:fault}a in Sec.~\ref{subsec:power}).
The nodal power injection is changed by \(\Delta P_i\) for a duration \(\Delta t_i\) and then restored to its pre-disturbance value:
\begin{equation}
P_i(t)
=
\begin{cases}
P_i^\ast+\Delta P_i, & 0\le t<\Delta t_i,\\
P_i^\ast, & t\ge\Delta t_i.
\end{cases}
\label{eq:pulse_disturbance}
\end{equation}
Here, \(P_i^\ast\) denotes the power injection of node \(i\) at the synchronized operating point, \(\Delta P_i\) the disturbance magnitude, and \(\Delta t_i\) its duration.
Each pair \((\Delta P_i,\Delta t_i)\) defines one power-disturbance scenario.

For each node, \(1{,}000\) scenarios were generated by log-uniform random sampling of the disturbance magnitude and duration.
The disturbance magnitude was sampled over
\begin{equation}
10^{-3}\sum_j K_{ij}
\le
|\Delta P_i|
\le
10^{2}\sum_j K_{ij},
\end{equation}
and the disturbance duration \(\Delta t_i\) was sampled log-uniformly from \(10^{-3}\) to \(100~\mathrm{s}\).
These sampling ranges define the initial disturbance ensemble and do not determine the validity range of the short-duration energy mapping.
For each scenario, the full swing equations are numerically integrated and the resulting trajectory is classified as secured or unsecured using the same phase-angle criterion applied to the state perturbations.
The network topology and nodal parameters are held fixed throughout the disturbance and subsequent recovery.

To relate the power disturbance to the perturbation energy defined in Sec.~\ref{subsec:methods_energy}, we approximate the target node as a single oscillator coupled to an equivalent reference bus (see Fig.~\ref{fig:fault}b in Sec.~\ref{subsec:power}).
The effective coupling strength of node \(i\) is represented by its weighted degree,
\begin{equation}
K_i
=
k_{w,i}
=
\sum_j K_{ij}.
\label{eq:effective_coupling}
\end{equation}
For the local linearized dynamics around the synchronized operating point, the corresponding restoring coefficient is
\begin{equation}
\kappa_i
=
\sum_j K_{ij}
\cos\left(\theta_i^\ast-\theta_j^\ast\right).
\label{eq:local_restoring}
\end{equation}
Defining the phase displacement from the synchronized operating point as
\begin{equation}
\delta_i
=
\theta_i-\theta_i^\ast,
\end{equation}
the local linearized dynamics are approximated by
\begin{equation}
M_i\ddot{\delta}_i
+
D_i\dot{\delta}_i
+
\kappa_i\delta_i
=
\Delta P_i.
\label{eq:linear_power_disturbance}
\end{equation}

For a sufficiently short disturbance, we consider the initial regime in which the phase displacement remains small and neglect the restoring term \(\kappa_i\delta_i\).
Over this interval, the additional power input is approximated as primarily changing the angular motion of the target node, yielding
\begin{equation}
M_i\dot{\omega}_i
+
D_i\omega_i
\approx
\Delta P_i,
\label{eq:reduced_power_disturbance}
\end{equation}
with \(\omega_i(0)=0\).
The angular velocity at the end of the pulse is then
\begin{equation}
\omega_i(\Delta t_i)
=
\frac{\Delta P_i}{D_i}
\left[
1-
\exp\left(
-\frac{D_i\Delta t_i}{M_i}
\right)
\right].
\label{eq:pulse_omega}
\end{equation}

The energy transferred to the angular motion of the target node during this initial regime is approximated by
\begin{equation}
E_i
=
\frac{1}{2}
M_i
\omega_i^2(\Delta t_i).
\end{equation}
Substituting Eq.~\ref{eq:pulse_omega} and solving for the disturbance magnitude gives
\begin{equation}
|\Delta P_i|
=
\sqrt{\frac{2E_i}{M_i}}
\,
\frac{D_i}
{
1-
\exp\left(
-D_i\Delta t_i/M_i
\right)
}.
\label{eq:power_duration_mapping}
\end{equation}
Substituting \(E_i=E_i^{\mathrm{S}}\) and \(E_i=E_i^{\mathrm{U}}\) into Eq.~\ref{eq:power_duration_mapping} yields the corresponding constant-energy contours in the \((\Delta P_i,\Delta t_i)\) plane (see Fig.~\ref{fig:fault}c in Sec.~\ref{subsec:power}).

Because Eq.~\ref{eq:power_duration_mapping} neglects the restoring term, the threshold comparison is restricted to the initial short-duration regime.
For each node, the damped local oscillation frequency is estimated as
\begin{equation}
\omega_{d,i}
=
\sqrt{
\frac{\kappa_i}{M_i}
-
\left(
\frac{D_i}{2M_i}
\right)^2
},
\label{eq:damped_local_frequency}
\end{equation}
and only disturbances satisfying
\begin{equation}
\Delta t_i
\le
\frac{\pi}{2\omega_{d,i}}
\label{eq:quarter_period}
\end{equation}
are included in the validation of the power-disturbance mapping.
This upper bound corresponds to the first quarter of the local damped oscillation period.
The full-network simulations retain the complete nonlinear swing dynamics, including coupling and damping; the approximation is used only to map the state-derived energy thresholds onto power--duration coordinates.
Consequently, only the subset of the \(1{,}000\) generated scenarios satisfying this short-duration condition contributes to the threshold-violation analysis for each node.

To quantify how consistently the mapped energy thresholds classify power-disturbance outcomes across nodes, we define an empirical threshold violation rate,
\begin{equation}
\epsilon_i
=
\frac{
N_i^{\mathrm{miss}}
}{
N_i^{\mathrm{match}}+N_i^{\mathrm{miss}}
},
\label{eq:threshold_violation}
\end{equation}
where \(N_i^{\mathrm{match}}\) denotes the number of sampled power-disturbance outcomes consistent with the mapped energy thresholds, and \(N_i^{\mathrm{miss}}\) denotes the number of sampled outcomes that violate them.
Specifically, an unsecured outcome below the \(E_i^{\mathrm{S}}\) contour or a secured outcome above the \(E_i^{\mathrm{U}}\) contour is counted as \(N_i^{\mathrm{miss}}\), whereas outcomes in the same two outer regions that satisfy the corresponding threshold condition are counted as \(N_i^{\mathrm{match}}\).
Disturbances lying between the two contours are excluded because both secured and unsecured outcomes are admissible within the mixed-outcome energy interval.
Accordingly, \(\epsilon_i\) represents the fraction of sampled outcomes outside the mixed-outcome interval that are inconsistent with the mapped thresholds.
It is therefore an empirical consistency measure with respect to the prescribed disturbance-sampling ensemble, rather than a probability weighted by the actual occurrence likelihood of disturbances.

\section*{Data availability}
The UK 54-node network dataset was obtained following the data-access information provided in Ref.~\cite{lee2024reinforcement}, and the Korean 193-node network dataset was obtained following Ref.~\cite{song2025kpg}.
The Swiss 161-node, UK 313-node, German 553-node, Spanish 838-node, and French 847-node network datasets were provided separately, with the corresponding sources cited as requested by the data providers~\cite{tyloo2019key,pagnier2019inertia}.
The source code for the secured-basin framework is publicly available at \url{https://github.com/jaiyonglee/secured-basin}.

\section*{Funding}
This research was supported by ANCHOR program through the Jeollanamdo Anchor Center, funded by the Ministry of Education(MOE) and the Jeollanamdo, Republic of Korea (2026-ANCHOR-14-009) (H.K.) and the National Research Foundation of Korea (NRF) under the grant Nos. RS-2026-25488703 (S.-W.S.), RS-2026-25468383 (S.H.L.), RS-2024-00341317 (M.J.L.), and RS-2022-NR071795 (H.K).


\section*{Declaration of competing interest}
The authors declare that they have no known competing financial interests or personal relationships that could have appeared to influence the work reported in this paper.

\section*{Acknowledgments}
We sincerely thank Cook Hyun Kim for insightful discussions that helped shape the energy-barrier concept.

\printcredits

\bibliographystyle{unsrt}

\bibliography{ref}

\end{document}